\documentclass[twocolumn]{aastex631}
\usepackage[utf8]{inputenc}
\usepackage[T1]{fontenc}
\usepackage{graphicx} 
\usepackage{amsmath}
\usepackage[sort&compress]{natbib}
\usepackage{float}
\usepackage{hyperref}

\usepackage[varg]{txfonts}
\usepackage{graphicx,color}
\usepackage{dcolumn}%
\usepackage{soul, ulem} 
\usepackage{hyperref,url}
\usepackage{xcolor}
\usepackage{array}
\usepackage{comment}
\usepackage{threeparttablex, tablefootnote}

\newcommand{\degs}[1]{#1^{\circ}}
\newcommand{\mev}{\;\text{MeV}}
\newcommand{\gev}{\;\text{GeV}}
\newcommand{\gray}{$\gamma$-ray\;}
\newcommand{\fermi}{\textit{Fermi}} 
\newcommand{\ts}{\mathrm{TS}}
\newcommand{\LL}{L_\gamma - L_{C,5\textrm{GHz}}}

\newcommand{\lcore}{L_{\textit{\tiny{C, 5GHz}}}}

\begin{document}

\title{The gamma-ray emission from Radio Galaxies \\
and their contribution to the Isotropic Gamma-ray Background}

\author{Antonio Circiello}
\affiliation{Department of Physics and Astronomy, 
Clemson University, 
Clemson, SC 29631, USA}
\author{Alex McDaniel}
\affiliation{Department of Physics and Astronomy, 
Clemson University, 
Clemson, SC 29631, USA}
\author{Mattia Di Mauro}
\affiliation{Istituto Nazionale di Fisica Nucleare, Sezione di Torino, 
Via P. Giuria 1, 10125 Torino, Italy}
\author{Christopher Karwin}
\affiliation{NASA Postdoctoral Program Fellow, 
NASA Goddard Space Flight Center, 
Greenbelt, MD, 20771, USA}
\author{Nikita S. Khatiya}
\affiliation{Department of Physics and Astronomy, 
Clemson University, 
Clemson, SC 29631, USA}
\author[0000-0002-6584-1703]{Marco Ajello}
\affiliation{Department of Physics and Astronomy, 
Clemson University, 
Clemson, SC 29631, USA}
\author{Fiorenza Donato}
\affiliation{Istituto Nazionale di Fisica Nucleare, Sezione di Torino, 
Via P. Giuria 1, 10125 Torino, Italy}
\author{Dieter Hartmann}
\affiliation{Department of Physics and Astronomy, 
Clemson University, 
Clemson, SC 29631, USA}
\author{Andrew W. Strong}
\affiliation{Max Plank Institut für extraterrestriche Physik, D-85748 Garching, Germany}

\begin{abstract}
    We evaluate the contribution to the Isotropic \gray Background (IGRB) coming from Radio Galaxies (RGs), the subclass of radio-loud Active Galactic Nuclei (AGN) with the highest misalignment from the line of sight (l.o.s.).
    Since only a small number of RGs are detected in $\gamma$ rays compared to the largest known radio population, the correlation between radio and \gray emission serves a crucial tool to characterize the \gray properties of these sources.
    We analyse the population of RGs using two samples.
    The first sample contains 26 sources individually detected by the Large Area Telescope (LAT) on board the \textit{Fermi Gamma-ray Space Telescope} at $\gamma$ rays.
    The second sample contains 210 RGs for which the \gray emission is not significantly detected by the LAT. 
    We use a stacking analysis to characterize the average properties of the \gray emission of the two samples, separately at first and then combined. 
    We then evaluate the correlation between their \gray emission and the emission from their radio core at 5 GHz, and we use it to determine their contribution to the IGRB.
    Due to the limited number of RGs detected at the $\gamma$-rays, information on the $\gamma$-ray luminosity function is limited. The $\LL$ correlation allows us to characterize it starting from the luminosity function of the radio cores, which is modeled with greater accuracy due to the larger number of sources detected at these frequencies.
    We find that the diffuse emission as extrapolated from the properties of the subthreshold RGs is lower than the one inferred from detected RGs, showing that the contribution of the population of RGs to the IGRB is lower than the previous estimates and it is around the $\sim$30\% level of the IGRB intensity.
\end{abstract}


\section{Introduction}\label{sec:intro}
\noindent Observations at high Galactic latitude of the \gray sky with the Large Area Telescope (LAT) on board the \textit{Fermi Gamma-ray Space Telescope} allowed for accurate measurements of the Isotropic Gamma-Ray Background (IGRB) spectrum from $100 \mev$ to $820 \gev$ \citep[][]{abdo+2010, Ackermann2015}.
The nature of this diffuse emission, however, is still a matter of debate. 
Several studies have tried to understand the composition of the IGRB, which has been speculated to originate from a large number of unresolved \gray sources \citep{Dermer+2007,Inoue+2011, DiMauro+2014, Ajello2015, Fermi+2015_igrbdm,Fornasa+15, Ajello+2020a, Roth+21, Fukazawa+2022, Korsmeier+22}.
A sizeable fraction of the IGRB intensity can be ascribed to the emission from the radio-loud Active Galactic Nuclei \citep[AGN; for a review, see][and references therein]{Padovani+2017} population.
Radio-loud AGN are a broad class of sources, whose observable properties strongly depend on the inclination of the jet with respect to the line of sight (l.o.s.).
It is therefore common to break it down into separate sub-classes, depending on this angle.
Sources with their jet axis within a few degrees from the l.o.s. (i.e., $<\degs{5}$) are referred to as blazars.
They represent the brightest subclass of radio-loud AGN, and their contribution to the IGRB has been extensively studied, reaching reasonably accurate constraints \citep{DiMauro:2013zfa,Ajello2015, DiMauro2018, Marcotulli+2020}.
The remaining sources, with higher separation between their jet axis and the l.o.s., are identified as misaligned AGN (MAGN).
The subclass of MAGN with highest jet misalignment from the l.o.s., known as Radio Galaxies (RGs), contains sources that are generally less luminous than blazars, but more numerous, and could therefore provide a significant contribution to the IGRB.
RGs are further divided into two separate categories by the Fanaroff-Riley classification \citep{FR+1974}, depending on their luminosity profile.
Fanaroff-Riley type I (FRI) sources display diffuse low luminosity radio emission.
Fanaroff-Riley type II (FRII) sources are usually brighter, with the emission peaking at the edges of their radio-lobes.
Previous works have investigated the RGs contribution to the IGRB using the limited amount of sources that the \fermi\;satellite was able to detect at \gray energies. 
These studies have shown that a robust portion of the IGRB might be ascribed to RGs, with \cite{DiMauro+2014} reporting that they are responsible for $\sim 10 - 83 \%$ of the IGRB, and \cite{Hooper+2016} evaluating their contribution to be $\sim$ 68-95\%. 
As the number of RGs detected by \fermi-LAT increased, \cite{Stecker+2019} suggested that their contribution might be significantly lower.\\
However, the RGs detected by \fermi-LAT are but a small portion of the population, as indicated by observations at radio frequencies \citep{YuanWang2012, Stecker+2019}.
While the vast majority of the RGs are too faint at $\gamma$ rays to be detected individually, the cumulative flux from the entire population might give a non-negligible contribution to the background.
A similar effect has been observed and well documented in blazars, where the contribution from the unresolved population has been shown to explain at least $10\%$ of the measured IGRB \citep{Stecker+1993, Padovani+1993a, Salamon+1994, abdo+2010, Abazajian+2011, Marcotulli+2020}.
Increasing the size of the sample through the inclusion of sources that are not detected individually by the \fermi-LAT can lead to a more realistic characterization of the population of RGs, and, in order, to a better constrained estimation of their contribution to the IGRB.
In this work we characterize the diffuse \gray emission from the RGs detected individually by the \textit{Fermi}-LAT and from the ones that are only detected in the radio band.
We use a stacking analysis to characterize the correlation between the radio and \gray emission of the two samples, at first separately, then combining the results to represent the entire population.
We use this correlation to characterize the contribution of RGs to the $\gamma$-ray background emission through their properties in the radio-band, which are known with higher accuracy.\\
The paper is organized as follows.
In $\S$ \ref{sec:data_sample} we lay out the samples used in the analysis.
In $\S$ \ref{sec:analysis} we present the \fermi-LAT data selection and analysis procedure.
In $\S$ \ref{sec:gamma_radio_corr} we discuss the correlation between the radio and $\gamma$-ray emissions from RGs, and the results for the two samples (separately and combined).
In $\S$ \ref{sec:diff_flux} we compute the contribution to the IGRB from the RG population.
In $\S$ \ref{sec:scd} we use the source count distribution for RGs to further test the results.
In $\S$ \ref{sec:disc} we summarize the results of this work.
Throughout the analysis we assume a standard flat $\Lambda$CDM cosmology, with Hubble constant $H_0 = 71\, \textrm{km s}^{-1} \textrm{Mpc}^{-1}$ and density parameters for the matter and dark energy components, respectively $\Omega_{M,0} = 0.27$ and $\Omega_{\Lambda,0} = 0.73$.

\begin{deluxetable*}{ccccccccccc}[b]
\label{tab:detected}
\tablecaption{List of RGs in the detected sample, with the integrated $\gamma$-ray flux ($log_{10}(F_\gamma)$) and spectral index ($\Gamma$) evaluated for each source.
The sources marked with $\dagger$ are modeled with a log-parabola $\gamma$-ray spectrum. For these sources, $\Gamma$ refers to the alpha index of the spectrum, while the beta index is kept fixed to its nominal value listed in the 4FGL-DR3.
Where available, redshift independent measurements for the distance are reported, taken from the NASA Extragalactic Database (NED). These distances are preferred to obtain the $\gamma$-ray luminosity ($L_\gamma$) instead of the luminosity distances evaluated from the redshift for nearby sources. We also report the detection significance from our analysis ($\sigma$, see \S \ref{sec:analysis}) and the Core Dominance parameter (R, see \S \ref{sec:stat_test}).\\}
\tablehead{
\colhead{Source} & \colhead{RA} & \colhead{Dec} & \colhead{Redshift} & \colhead{Distance}
& \colhead{$\log_{10}(L_{C,5\textrm{GHz}})$} & \colhead{$\log_{10}(L_\gamma)$} & \colhead{$\log_{10}(F_\gamma)$} &
\colhead{$\Gamma$} & $\sigma$ & \colhead{R}\\
\colhead{} & \colhead{[$\degs{}$]} & \colhead{[$\degs{}$]} & \colhead{-}  & \colhead{[Mpc]} & \colhead{[erg/s]} & \colhead{[erg/s]} & \colhead{[$\#/\mathrm{cm}^2/\mathrm{s}$]} & \colhead{-} & \colhead{-} & \colhead{-}}
\startdata
NGC 315 & 14.45 & 30.40 & 0.016 & 68 & 40.21 & $41.9^{+0.4}_{-0.2}$ & $-9.6_{-0.2}^{+0.2}$ & $2.3_{-0.4}^{+0.4}$ & 9.6 & 0.98\\ 
TXS 0149+710 & 28.36 & 71.24 & 0.022 & - & 40.19 & $42.5^{+0.3}_{-0.2}$ & $-9.6_{-0.2}^{+0.2}$ & $1.9_{-0.2}^{+0.2}$ & 11.3 & 7.97 \\
B3 0309+411B & 48.24 & 41.33 & 0.136 & - & 41.99 & $43.6^{+0.6}_{-0.2}$ & $-9.8_{-0.3}^{+0.3}$ & $2.3_{-0.4}^{+0.5}$ & 5.3 & 0.06\\ 
4C +39.12 & 53.58 & 39.34 & 0.021 & - & 39.84 & $42.5^{+0.5}_{-0.2}$ &$-9.6_{-0.2}^{+0.2}$ &  $1.9_{-0.3}^{+0.2}$ & 11.1 & 1.46\\ 
Pictor A & 79.91 & -45.74 & 0.035 & - & 41.14 & $42.7^{+0.1}_{-0.2}$ & $-9.6_{-0.3}^{+0.1}$  &  $2.2_{-0.2}^{+0.4}$ & 10.0 & 0.06\\ 
NGC 2329 & 107.25 & 48.66 & 0.019 & 70.2 & 39.31 & $42.2^{+0.3}_{-0.1}$ & $-10.0_{-0.3}^{+0.2}$  &  $1.6_{-0.2}^{+0.3}$ & 9.2 & 0.51\\ 
NGC 2484 & 119.69 & 37.78 & 0.043 & 171 & 40.53 & $42.5^{+0.1}_{-0.2}$ & $-9.6_{-0.3}^{+0.1}$  &  $2.8_{-0.5}^{+0.7}$ & 7.6 & 0.79\\ 
NGC 3078 & 149.59 & -26.93 & 0.009 & 36 & 38.94 & $41.2^{+0.4}_{-0.2}$ &$-9.7_{-0.3}^{+0.2}$  &  $2.3_{-0.4}^{+0.5}$ & 7.3 & 2.57\\ 
B2 1113+29 & 169.16 & 29.26 & 0.047 & - & 40.01 & $42.6^{+0.9}_{-0.1}$ & $-10.4_{-0.6}^{+0.3}$  &  $1.7_{-0.5}^{+0.7}$ & 4.6 & 0.22 \\ 
3C 264 & 176.24 & 19.63 & 0.022 & 113 & 40.18 & $42.7^{+0.3}_{-0.2}$ & $-9.5_{-0.2}^{+0.1}$  &  $2.0_{-0.3}^{+0.2}$  & 14.6 & 0.05 \\ 
NGC 3894 & 177.25 & 59.42 & 0.011 & 45.6 & 39.89 & $41.5^{+0.3}_{-0.2}$ & $-9.7_{-0.2}^{+0.2}$  &  $2.2_{-0.3}^{+0.4}$  & 10.5 & 1.02\\ 
NGC 4261 & 184.91 & 5.84 & 0.007 & 32.4 & 38.86 & $41.2^{+0.5}_{-0.2}$ & $-9.7_{-0.2}^{+0.2}$  &  $2.2_{-0.4}^{+0.4}$  & 8.8 & 0.01 \\ 
M $87^{\dag}$  & 187.71 & 12.39 & 0.004 & 15.2 & 39.74 & $44.5^{+0.1}_{-0.2}$ & $-8.8_{-0.1}^{+0.1}$  &  $2.1_{-0.2}^{+0.1}$  & 44.0 & 0.06\\ 
PKS 1234-723 & 189.24 & -72.53 & 0.024 & - & 39.98 & $42.2^{+0.4}_{-0.2}$ & $-9.6_{-0.3}^{+0.2}$  &  $2.4_{-0.4}^{+0.4}$  & 6.6 & 0.48\\ 
TXS 1303+114 & 196.60 & 11.23 & 0.086 & 352 & 41.17 & $43.4^{+0.4}_{-0.1}$ & $-9.8_{-0.3}^{+0.2}$  &  $2.0_{-0.3}^{+0.4}$  & 7.5 & 2.69\\ 
PKS $1304-215^{\dag}$ & 196.70 & -21.81 & 0.126 & - & 41.31 & $43.1^{+0.4}_{-0.2}$ & $-9.3_{-0.1}^{+0.2}$  &  $2.2_{-0.3}^{+0.3}$  & 14.6 & 1.67\\ 
3C 303 & 220.78 & 52.03 & 0.141 & - & 41.67 & $43.8^{+0.4}_{-0.2}$ & $-9.8_{-0.2}^{+0.2}$  &  $2.1_{-0.3}^{+0.4}$  & 10.2 & 0.60\\ 
B2 1447+27 & 222.40 & 27.77 & 0.031 & - & 39.54 & $42.9^{+0.3}_{-0.1}$ & $-10.0_{-0.2}^{+0.2}$  &  $1.5_{-0.2}^{+0.3}$ & 10.7 & 1.71 \\
PKS 1514+00 & 229.14 & 0.27 & 0.052 & - & 40.61 & $42.9^{+0.2}_{-0.3}$ & $-9.5_{-0.2}^{+0.2}$  &  $2.7_{-0.4}^{+0.6}$  & 8.2 & 0.45\\ 
TXS 1516+064 & 229.65 & 6.24 & 0.102 & 441 & 41.07 & $43.7^{+0.5}_{-0.1}$ & $-9.9_{-0.3}^{+0.2}$  &  $1.8_{-0.3}^{+0.4}$  & 7.5 & 1.89\\ 
PKS B1518+045 & 230.29 & 4.36 & 0.052 & 213 & 40.54 & $42.7^{+0.4}_{-0.2}$ & $-9.7_{-0.3}^{+0.2}$  &  $2.4_{-0.4}^{+0.6}$  & 6.2 & 5.86\\
NGC $6251^{\dag}$ & 247.67 & 82.57 & 0.024 & 97.6 & 40.14 & $43.6^{+0.3}_{-0.5}$ & $-8.9_{-0.1}^{+0.1}$ &  $2.1_{-0.1}^{+0.1}$  & 42.8 & 0.46\\ 
NGC 6454 & 266.24 & 55.70 & 0.030 & - & 40.41 & $42.2^{+0.3}_{-0.2}$ & $-9.9_{-0.2}^{+0.2}$ & $2.2_{-0.3}^{+0.6}$ & 6.8 & 4.62 \\
PKS 1839-48 & 280.87 & -48.59 & 0.111 & - & 41.39 & $43.7^{+0.4}_{-0.2}$ & $-9.7_{-0.2}^{+0.2}$ &  $2.1_{-0.3}^{+0.4}$  & 8.0 & 0.32\\ 
PKS 2225-308 & 336.98 & -30.52 & 0.056 & - & 40.04 & $43.2^{+0.5}_{-0.1}$ & $-9.9_{-0.3}^{+0.2}$ & $1.8_{-0.3}^{+0.4}$  & 8.3 & 0.27\\ 
PKS 2300-18 & 345.72 & -18.70 & 0.129 & - & 41.95 & $43.7^{+0.3}_{-0.2}$ &$-9.7_{-0.3}^{+0.2}$  &  $2.2_{-0.3}^{+0.4}$ & 8.1 & 2.10\\ 
\enddata
\end{deluxetable*}

\begin{deluxetable}{cccccc}[b]
\label{tab:undetected}
\tablecaption{List of RGs in the undetected sample. For each source, we list the equatorial coordinates, the redshift, the logarithm of the luminosity of the radio core at 5 GHz ($\log_{10}(L_{C, 5GHz})$), and the logarithm of the core dominance parameter ($\log_{10}(R)$). The full table is available in machine readable format as supplemental material.}
\tablehead{\colhead{Source} & \colhead{RA} & \colhead{Dec} & \colhead{Redshift} & \colhead{$\log_{10}(L_{C, 5GHz})$} & \colhead{$\log_{10}(R)$}\\
\colhead{[-]} & \colhead{[$\degs{}$]} & \colhead{[$\degs{}$]} & \colhead{[-]} & \colhead{[erg/s]} & \colhead{[-]}}
\startdata
4C12.03 & 2.47  & 12.73 & 0.156 & 40.07 & -3.15\\ 
3C16    & 9.44  & 13.33 & 0.405 & 39.92 & -4.47\\ 
3C19    & 10.23 & 33.17 & 0.482 & 40.36 & -4.39\\ 
3C20    & 10.79 & 52.06 & 0.174 & 40.05 & -4.07\\ 
3C31    & 16.85 & 32.41 & 0.017 & 39.45 & -1.96\\
...     & ...     & ...     & ...     & ...     & ... \\
\enddata
\end{deluxetable}

\section{Radio galaxy samples}\label{sec:data_sample}

\begin{figure}[h]
    \centering
    \includegraphics[width = \linewidth]{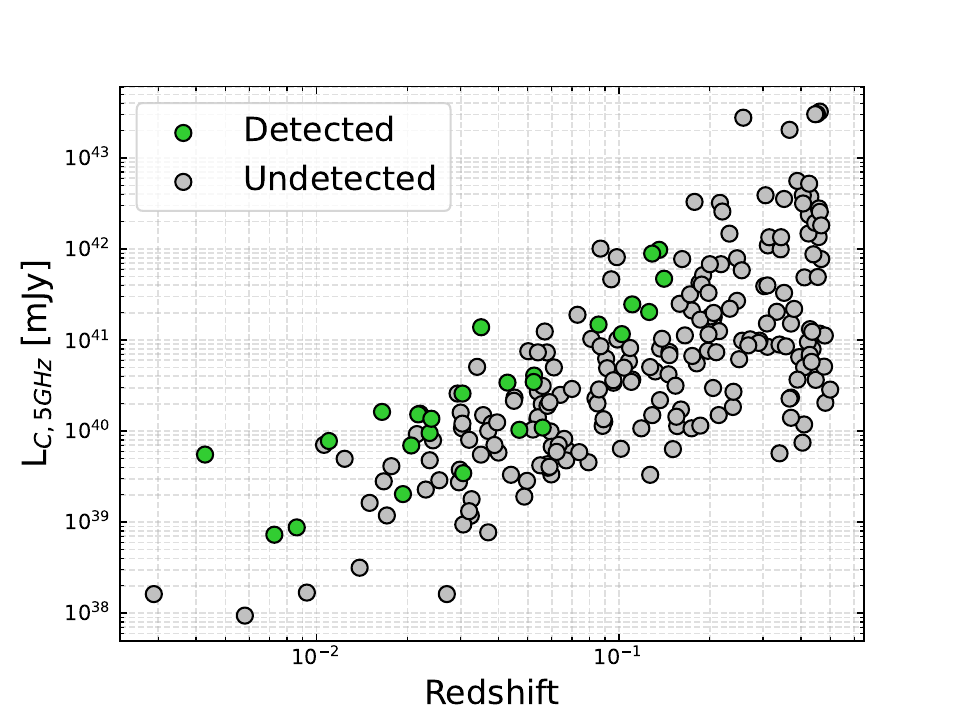}
    \caption{5 GHz core luminosity versus redshift for the sample of detected RGs at $\gamma$-rays (green circles) and for the sample of subthreshold RGs (gray circles) presented in Sect. \ref{sec:data_sample}.}
    \label{fig:sample}
\end{figure}

\noindent We consider two separate RG samples for the analysis, one containing only sources that are individually detected in both the \gray and radio band, the other, of much larger size, containing sources only detected in the radio band, whose \gray flux falls under the detection threshold for inclusion in \textit{Fermi}-LAT catalogs.\footnote{The detection threshold is TS > 25 (see \S~\ref{sec:analysis} for the definition of TS).}
The first sample comprises 26 FRI and FRII galaxies from the 4FGL-DR3 \citep{4fgldr3} catalog.
From the 45 sources classified as FRI or FRII in the catalog, we exclude sources following similar criteria to \cite{Khatiya+23}. 
We remove the sources detected by the LAT as extended (Fornax A and the Centaurus A lobes), the sources with negligible emission between 1 GeV and 800 GeV (3C 17, 3C 111, PKS 2153-69), the sources with no conclusive
data available (i.e., sources that only have upper limits) for the core emission at 5 GHz (PKS 0235+017, NGC 6328, PKS 2324-02, PKS 469 2327-215, PKS 2338-295), and Centaurus B as it lies on the Galactic plane.
Additionally, we remove the sources that have less than 1\% probability to be steady sources according to their variability index\footnote{The variability index is defined in the 4FGL as the sum of the logarithmic likelihood difference between the flux fitted in each time interval and the average flux over the full catalog interval. In the DR3, a value over 24.725 in 12 intervals indicates <1\% chance of the source being steady.}(IC 1531, NGC 1218, IC 310, 3C 120, PKS 0625-35, NGC 2892, Centaurus A).
The sample of detected sources, with the relevant data, is listed in Table \ref{tab:detected}.
For the subthreshold sample, we start from the 1103 radio loud AGNs listed in \citet{YuanWang2012}.
From this list we exclude the sources that have no conclusive data for the radio luminosity of the core at 5 GHz, which leaves 566 viable sources.
Furthermore, we exclude sources that fall within the $95\%$ confidence radius of a 4FGL source or that are within $\degs{0.1}$ from a source in the ROMA-BZCat catalog \citep{bzcat}, which remove 67 and 63 additional sources, respectively.
These extra cuts remove contamination from \gray sources that are associated with something other than a RG and potential contamination from blazars.
Finally, we remove all the sources with redshift above 0.5.
At this redshift even the brightest detected RGs are well below the threshold for detection, and our analysis can not distinguish them from other background effects.
After the cuts, 210 RGs are left in the sample, listed in Table \ref{tab:undetected}.
The distribution of redshift vs. core luminosity for both the \gray detected and undetected samples can be seen in Fig.~\ref{fig:sample}.  
The sample of detected RGs is concentrated at low redshifts, with $z \lesssim 0.15$, while the undetected RGs extend to higher values, up to $z = 0.5$, where we cut the sample. 
Even though many undetected sources show a high luminosity in radio and, therefore, a high predicted luminosity at $\gamma$-rays, the increased distances make them difficult to observe at these energies.

\section{Fermi-LAT data selection\\ and analysis}\label{sec:analysis}
\noindent For each source, we analyse data collected by \fermi-LAT between  August 4, 2008 and June 21, 2024, for $15.9$ years.
We select events in the energy band between $1 \gev$ and $800 \gev$, and use the filters DATA\_QUAL $> 0$ and LAT\_CONFIG $== 1$ to define the Good Time Intervals (GTI) for the FermiPy (v1.2.0) analysis. 
The choice of selecting photons above 1 GeV is done to optimize the Point Spread Function (PSF) of the instrument and maximize the signal-to-noise ratio.
We use the Pass 8 \citep{pass8} Instrument Response Function (IRF) P8R3\_SOURCE\_V3.
We define a $\degs{10}\times\degs{10}$ region of interest (ROI) 
centered on the source, with a pixel size of $\degs{0.08}$. The size of the ROI, as well as the pixel size and the energy interval, are chosen following the optimal choices for these parameters found in previous studies that employed the same stacking analysis method (e.g., \citealt{Khatiya+23, McDaniel+2023, McDaniel+2023b}).
We include in the model the Galactic diffuse (gll\_iem\_v07), the isotropic diffuse (iso\_P8R3\_SOURCE\_v3\_v1), all the sources from the 4FGL-DR3 catalog \citep{4fgldr3} within a radius of $\degs{15}$ from the target, and a point source with a power law spectrum to represent our target source\footnote{At energies above 1 GeV this is a good approximation even for a source modeled with a curved spectral shape.}, if it is part of the subthreshold sample.

Each target in the detected sample is analyzed in two steps to estimate its spectral parameters.
In the first step, we optimize the model leaving the normalization and photon index of the Galactic diffuse, the normalization of the isotropic diffuse,
the parameters of all the sources within $\degs{7}$ of the target with $\ts>16$, and the normalization and photon index of the target source\footnote{or the normalization and alpha index, if the target has a LogParabola spectrum in the 4FGL-DR3.} as free parameters.
Additionally, we use the \textit{find\_sources()} method to scan the ROI for new sources and add them to the model.
In the second step, we fix all the parameters in the optimized models, except the normalization and photon index of the Galactic diffuse emission and the normalization of the isotropic diffuse emission, 
and scan over a range of values for the integrated photon flux and photon index of the target source, assigning a TS value to each possible pair, defined as:
\begin{equation}
    \ts = - 2 \ln\left(\frac{L_0}{L}\right)
\end{equation}
where $L$ is the likelihood of the model when the source is assigned the respective values of photon index and normalization, and $L_0$ is the likelihood for the null hypothesis (i.e. the same model without the target source).
This iteration produces a TS profile for each source, that peaks at the best-fit values for the parameters of interest.
For sources in the detected sample, the TS profile peaks at $TS>25$, which corresponds approximately to a significance $\sim4.6\sigma$ for 2 degrees of freedom (d.o.f.).
The results of this analysis are listed in Table \ref{tab:detected}, in which we report the \gray flux and index obtained for each source of the detected sample. 
Note that M 87, PKS 1304-215 and NGC 6251 are described by a log-parabola spectrum in the 4FGL-DR3 and are therefore modeled as such in our analysis. For these sources the photon indexes reported in Table \ref{tab:detected} are the alpha indexes of the spectrum, while the beta indexes are kept fixed to their nominal values listed in the catalog.

\begin{figure}
    \centering
    \includegraphics[width = \linewidth]{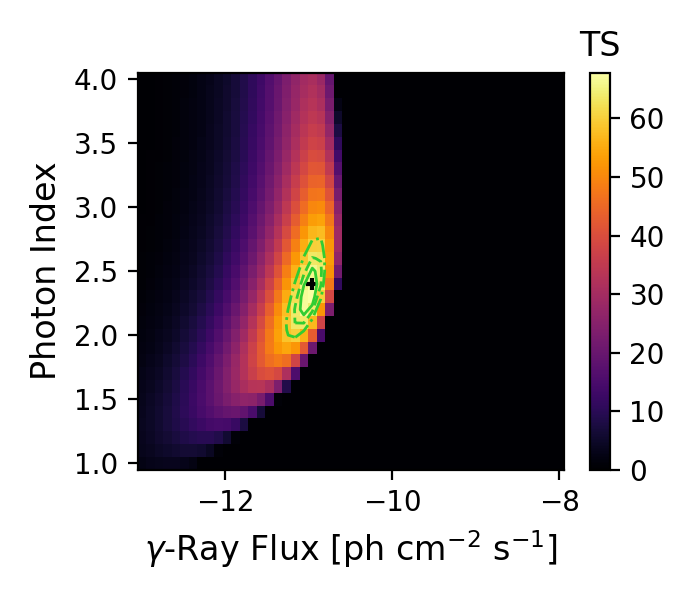}
    \caption{Stacked TS profile for the subthreshold RG sample. The coordinates of the black cross indicate the best estimates for the average integrated flux and photon index of the population. The green contours are the uncertainties at the $1\sigma$, $2\sigma$, and $3\sigma$ levels respectively.}
    \label{fig:TS_profile}
\end{figure}

The analysis of targets in the subthreshold sample follows the same two steps.
The peak TS for these targets individually, however, is too low (i.e., $\mathrm{TS}<25$) to indicate a significant detection.
Therefore we apply a stacking technique wherein the individual TS profiles are summed in order to characterize the average spectral energy density (SED) of the sample (Fig.~\ref{fig:TS_profile}).
This method has been successfully employed to obtain upper limits on dark matter interactions \citep{Ackermann+2011, McDaniel+2023b, Circiello+2025}, for the detection of the extragalactic background light \citep{Abdollahi+2018}, and the study of extreme blazars \citep{Paliya+2019}, star-forming galaxies \citep{Ajello+2020a}, fast black-hole winds \citep{Ajello+2021}, and molecular outflows \citep{McDaniel+2023}.
As shown in Fig.~\ref{fig:TS_profile}, the average emission of subthreshold RGs is detected significantly at the $7.9 \sigma$ level for 2 d.o.f., assuming that asymptotic behavior of Chernoff’s theorem applies\footnote{In case of asymptotic behavior of Chernoff’s theorem \citep{Chernoff+1952}, the $TS$ distribution of the null hypothesis follows a Poissonian distribution yielding a distribution compatible with a $\chi^2/2$ distribution for two degrees of freedom.}.
The average flux is $F_\gamma = 1.1^{+1.4}_{-0.3} \times 10^{-11} \textrm{ ph cm}^{-2}\textrm{ s}^{-1}$, and the average photon index is $\Gamma = 2.4^{+0.2}_{-0.2}$.

\section{Radio-Gamma Correlation}\label{sec:gamma_radio_corr}

\noindent Due to the limited number of RGs detected by \textit{Fermi}-LAT with a $\ts>25$, the \gray properties of the population are not known with great accuracy.
This implies that we cannot estimate precisely the contribution of RGs to the IGRB using only the detected sources by the LAT.
However, the \gray and radio emissions from these sources are correlated by a rather simple empirical relation, which has been proven to hold in various studies \citep[see][]{Inoue+2011,DiMauro+2014, Hooper+2016, Stecker+2019}.
Following the previous determinations, we adopt a linear correlation between the logarithms of the core luminosity at $5$ GHz and the \gray luminosity between 1 GeV and 800 GeV:
\begin{equation}
    \log_{10}\left(\frac{L_\gamma}{[\textrm{erg/s]}}\right) = \alpha \log_{10} \left(\frac{\lcore}{L_{*}}\right) + \beta
\end{equation}
where $L_*$ is a normalization factor chosen to roughly correspond to the median of the core luminosities; in our case  $L_* = 10^{42} \textrm{ erg s}^{-1}$.
The values of $\lcore$ for sources in the detected sample are taken from \citet{Khatiya+23}, while the ones for sources in the undetected sample are from \citet{YuanWang2012}.
We evaluate the correlation separately for each sample, employing a stacking procedure, this time in the $\alpha$-$\beta$ space.
To do this we convert the flux-index TS profile for each RG in the sample of interest to the $\alpha$-$\beta$ space using the RG distance and the average photon index of the sample ($\Gamma \sim 2.2$ for the detected RGs and $\Gamma \sim 2.4$ for the undetected RGs). 
As previously, we then sum the individual TS arrays to obtain the best-fit parameters.
For the detected sample we find $\alpha = 1.0^{+0.1}_{-0.1}$ and $\beta = 44.0^{+0.1}_{-0.1}$, with a peak TS of $TS = 3488$, which corresponds to a confidence level of $\sim 59.0 \sigma$ (for 2 d.o.f.).
For the subthreshold sample we find $\alpha = 0.6^{+0.1}_{-0.1}$ and $\beta = 42.6^{+0.1}_{-0.2}$, with a peak TS of $TS = 87.83$, which corresponds to a confidence level of $\sim 9.1 \sigma$ (for 2 d.o.f.).

\begin{figure}
    \centering
    \includegraphics[width = \linewidth]{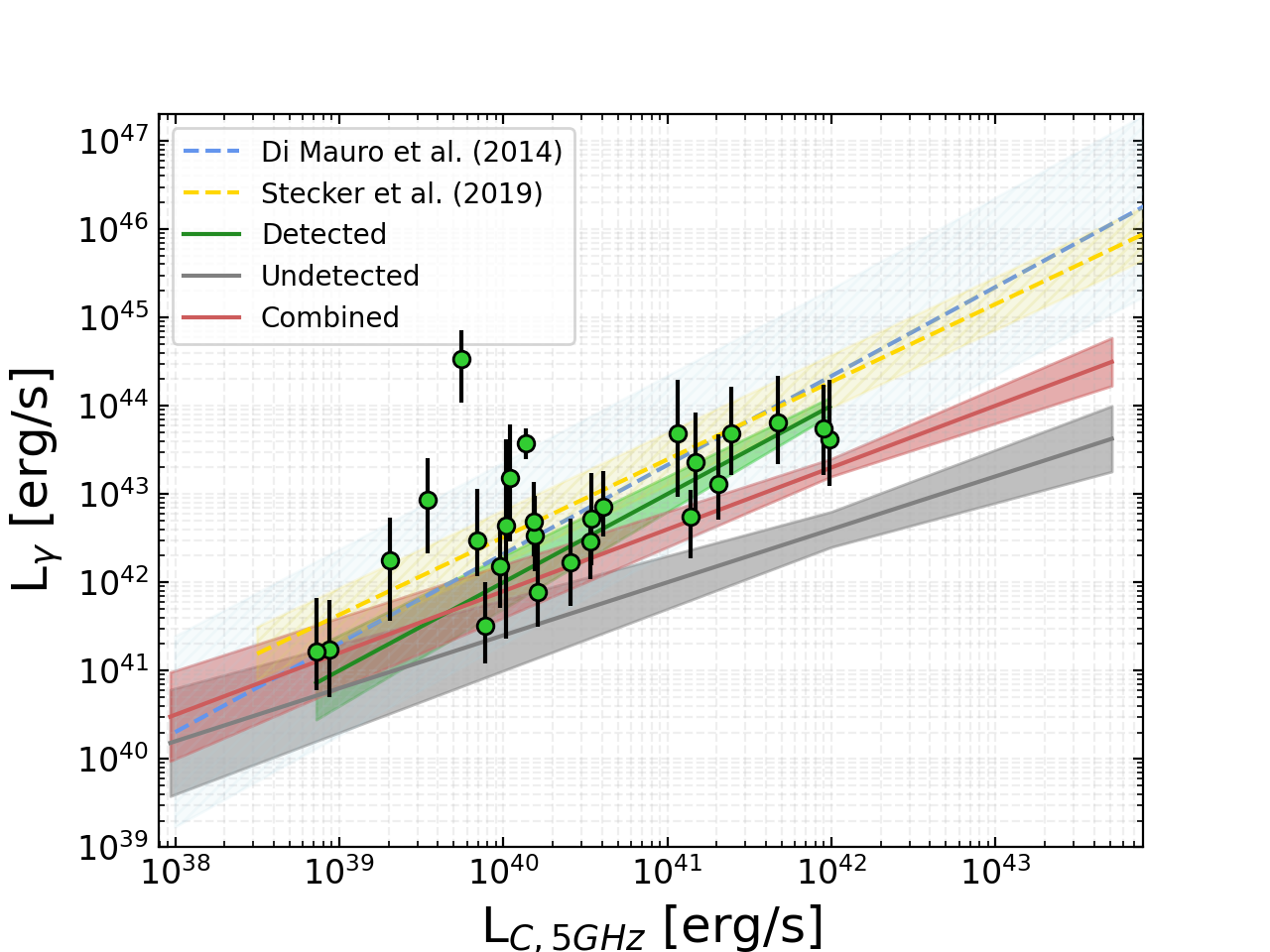}
    \caption{Correlation between the 1 -- 800 GeV \gray luminosity and the luminosity of the radio core at 5 GHz for the RGs analysed in this paper.
    The green points are the individually detected RGs. The values for the \gray luminosity are obtained through their individual TS profiles obtained in this analysis, while the core luminosities come from \cite{Khatiya+23}.
    The green band is the $L_\gamma - L_{C,5\textrm{GHz}}$ correlation for the detected RGs with its $1\sigma$ uncertainty.
    The gray band is the  $1\sigma$ band on the $L_\gamma - L_{C,5\textrm{GHz}}$ correlation derived from the analysis of the  subthreshold sample.
    The red band represents the $L_\gamma - L_{C,5\textrm{GHz}}$ correlation from the combination of the detected and undetected samples, with its $1\sigma$ uncertainty.
    We also include the results from \citet{DiMauro+2014} and \citet{Stecker+2019}, the dashed blue and yellow lines, respectively, with their $1\sigma$ uncertainties. 
    }
    \label{fig:correlation}
\end{figure}

The $L_\gamma - L_{C,5\textrm{GHz}}$ correlation for each sample is shown in Fig.~\ref{fig:correlation}.
In the same figure are displayed the data points for the individually detected RGs. 
It is apparent from Fig.~\ref{fig:correlation} that sub-threshold RGs result on average fainter than LAT detected RGs.
This could happen primarily for two reasons, a physical one and a spurious one.
The physical reason for the difference may lie in the fact that detected RGs are on average more beamed (likely because their inner jet is more aligned along our l.o.s.) than undetected ones. 
This would make detected RGs on average brighter than undetected ones.
On the other hand, the stacking procedure may dilute the signal of the undetected RGs by introducing fields with effectively no emission from RGs.
In \S~\ref{sec:stat_test} we perform tests about this scenario.

We also present the $L_\gamma - L_{C,5\textrm{GHz}}$ correlation obtained from the combination of the stacked profiles of the two samples. 
These results are a more realistic representation of the entire population of RGs, which takes into account the effects of the combined emission of RGs that are too faint to be detected individually, as well as the emission of the brightest RGs, which are detected by the \fermi-LAT.
We obtain $\alpha = 0.7^{+0.1}_{-0.1}$ and $\beta = 43.3^{+0.1}_{-0.1}$, with  a peak of $TS \sim 4058$, which corresponds to a confidence level of $\sim 63\sigma$ (for 2 d.o.f.).

\subsection{Test on the correlation}\label{sec:stat_test}

\noindent The correlation parameters for the luminosity of the \gray and radio emitting regions of the sources in the two samples differ significantly even though they are supposed to be part of the same class of sources.
This hints that $\alpha$ and $\beta$ may depend on other properties not accounted for in our analysis.
Examples of such parameters might be the viewing angle of the jet or the beaming factors of the radio and \gray jet, which can affect the ratio of the two luminosities since they are not isotropic emissions. 
Even though this effect is somewhat expected, it introduces a dependency of the results on the selection of the sample.
Because beaming factors and viewing angles are unfortunately not available for the vast majority of our sources, we try to rule out the unphysical explanation by performing some additional tests, as follows.

Due to their faintness, it is very likely that a large fraction of the stacked sources yield effectively no signal, and possibly bias the stacked TS profile towards lower values of the parameters.
For this reason, we test how the $L_\gamma - L_{C,5\textrm{GHz}}$ correlation changes if we stack only the RGs for which we expect a reasonably bright flux.
We start by defining the average point source sensitivity
for an individual source as $f_F = 10^{-12}\textrm{ erg s}^{-1}\textrm{ cm}^{-2}$ (in the 1-800 GeV band)\footnote{We use an average value, as the detection sensitivity depends on several parameters such as the energy range, location of the sky, source SED, and so on.
For a complete review of the top-level performances of the telescope, we refer to \url{https://s3df.slac.stanford.edu/data/fermi/groups/canda/lat_Performance.htm}}.
This value is taken as the minimum flux that includes $90\%$ of all 4FGL-DR3 sources at $|b| > \degs{10}$.
Then, we define the stacking sensitivity for a sample of N galaxies as the  $f_S = f_F/\sqrt{N}$.
We repeat the stacking only including the 100 sources with the highest predicted flux $>f_S$ (from the $L_\gamma - L_{C,5\textrm{GHz}}$ correlation of $\S$ \ref{sec:gamma_radio_corr})\footnote{This way, we lower the effective sensitivity by an order of magnitude, while making sure that the emission from all the selected RGs is within this adjusted threshold.}.
 The analysis of this sub-selection of sources did not yield any significant change in the parameters for the $L_\gamma - L_{C,5\textrm{GHz}}$ correlation, which keep the values $\alpha = 0.6^{+0.1}_{-0.1}$ and $\beta = 42.5^{+0.2}_{-0.1}$ evaluated from the full sample, with an only marginally lowered TS of  $71.44$, corresponding to a confidence level of $8.2\sigma$ (for 2 d.o.f.).
This result rules out the possibility that the lower luminosity observed on average for the sample of undetected RGs is produced by stacking "empty" positions in the sky. 

\begin{figure}[h]
    \centering
    \includegraphics[width = \linewidth]{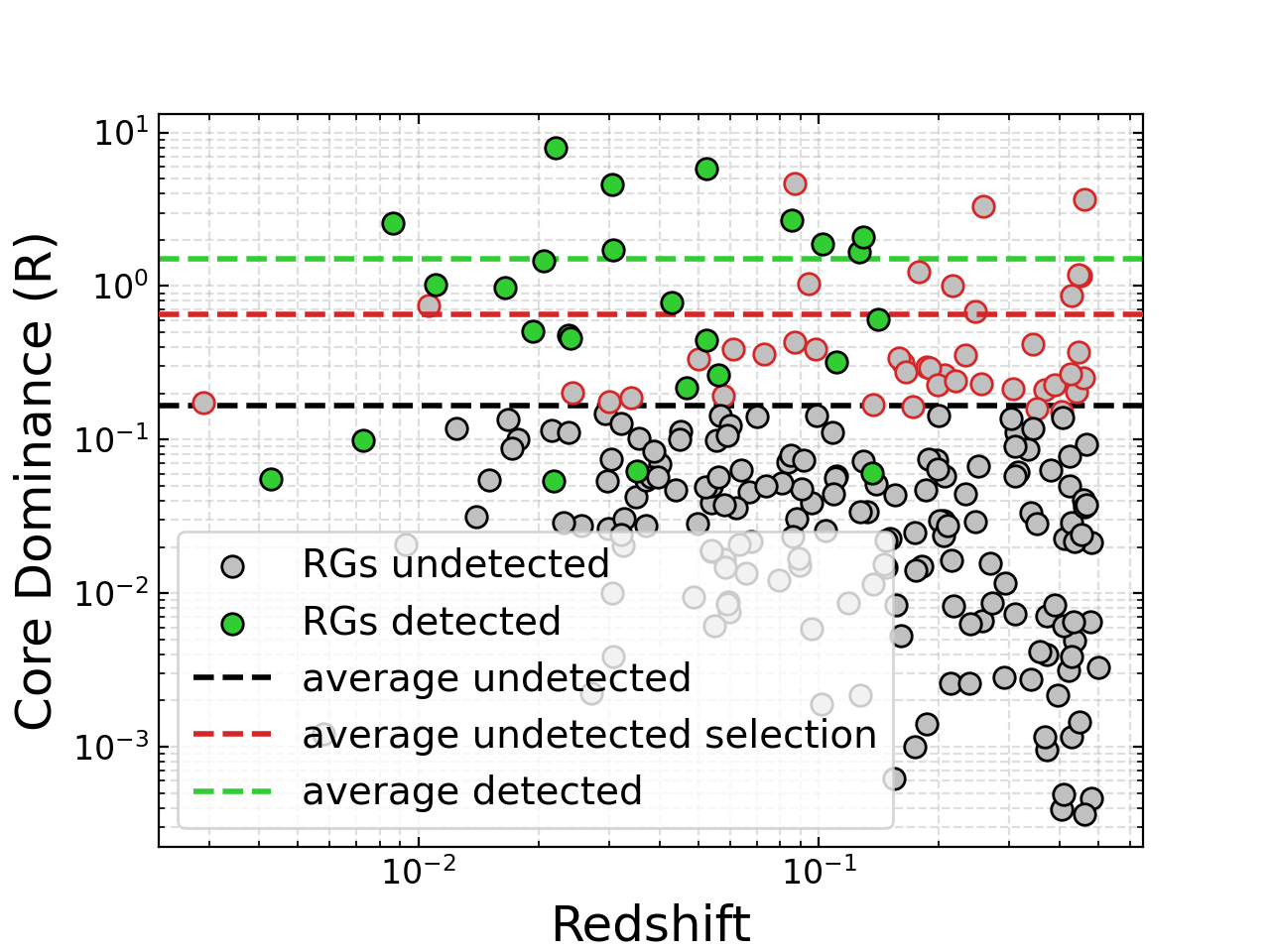}
    \caption{Core dominance (see eq. \ref{eq:core_dom}) versus redshift for the sample of detected RGs at $\gamma$-rays (green circles) and for the sample of subthreshold RGs (gray circles). The subsample of undetected RGs chosen to have R $>0.15$ \textbf{is highlighted with red contours}. The green, gray and red lines represent the average R of the detected RGs, the undetected RGs, and the subsample of undetected RGs, respectively. }
    \label{fig:core_dom}
\end{figure}

Having excluded the spurious nature of the difference in the $\LL$ correlation for the two samples, the parameters must depend on physical quantities that were not included in this analysis.
In principle, the main difference could be ascribed to the jet misalignment from the l.o.s., which strongly affects the observed luminosity of an AGN.
While data on the misalignment is not readily available for our sources, we can study the difference in orientation through the core dominance (R), defined as:
\begin{equation}
    \label{eq:core_dom}
    R = \frac{S_{C, 5GHz}}{S_{tot, \nu} - S_{C, 5GHz}} \cdot \left(\frac{\nu}{5 \mathrm{GHz}} \right)^{-\alpha_E} \cdot \left(1 + z \right)^{-\alpha_E}
\end{equation}
where $S_{C, 5GHz}$ is the flux from the radio core at $5$ GHz, while $S_{tot, \nu}$ is the total radio flux of the source, at frequency $\nu$.
The factor $(\nu/5\mathrm{ GHz})^{\alpha_E}$ rescales the total flux at 5 GHz.
The factor $(1+z)^{-\alpha_E}$ is the $k$-correction for the total radio flux, where $\alpha_E$ is the total radio index, assumed here to be $\alpha_E = 1$.
This factor is omitted for the core flux, as we assume $\alpha_C = 0$ \citep[see][]{Fan+2003}.
Fig.~\ref{fig:core_dom} shows the R parameter for sources in both samples.
On average, sources in the detected sample have a higher core dominance than sources in the undetected sample.
However, it is unclear whether or not this is the cause of the discrepancy in the level of emission between detected and undetected RGs.
While this could be tested by isolating a sample of undetected RGs with high R, only a few sources display a value of this parameter that is comparable to the detected RGs, which strongly affects the significance of the stacking analysis.
We selected RGs with R $> 0.15$, which is the highest possible cut on this sample to still have results with a significance of $\sim5\sigma$ (for 2 d.o.f.), though we obtain no change in the parameters $\alpha$ and $\beta$ of the $\LL$ correlation.

\section{Integrated Gamma-ray Emission of Radio Galaxies}\label{sec:diff_flux}
\noindent The determination of the contribution from the RG population to the IGRB requires knowledge of the \gray luminosity function (GLF) that describes it.
This is defined as the number of sources per comoving volume with luminosity in the range $\log L_{\gamma}$, $\log L_{\gamma} + d\log L_{\gamma}$
\begin{equation}
    \rho_{\gamma}(L_{\gamma}, z) = \frac{d^2 N}{dV d\log L_{\gamma}}.
\end{equation}
Due to the limited number of RGs observed at \gray energies, a direct determination of the GLF is not feasible.
However, RGs have been extensively characterized at radio frequencies, and their radio luminosity function (RLF) is known with reasonable accuracy \citep{Willot+2001, YuanWang2012, Yuan+2018}.
The same reasoning that led to the investigation of the correlation between \gray and radio emission in $\S$ \ref{sec:gamma_radio_corr} can be applied to relate the two luminosity functions, so that the GLF can be evaluated from the RLF as:
\begin{equation}
    \label{eq:GLF}
    \rho_\gamma (L_\gamma, z) = k \frac{d \log_{10}L_R}{d\log_{10} L_\gamma} \rho_R(L_R(L_\gamma), z)
\end{equation} 
where we assume $k=1$ (see, however, Sec. \ref{sec:scd}), while the derivative factor is obtained from the $L_\gamma - L_{C,5\textrm{GHz}}$ correlation presented in $\S$ \ref{sec:gamma_radio_corr}.
We consider the RLF presented in \citet{Yuan+2018} to describe the core emission at 5 GHz from the RG population.
This work presents a parametric core RLF in a double power-law form:
\begin{equation}
    \rho_c(L_c, z) = e_1(z) \phi_1 \left[ \left( \frac{L_c/e_2(z)}{L_*}\right)^\beta + \left(\frac{L_c/e_2(z)}{L_*} \right)^\gamma\right]^{-1}
\end{equation}
where
\begin{equation}
    e_1(z) = \frac{(1+z_c)^{p_1} + (1+z_c)^{p_2}}{\left(\frac{1+z_c}{1+z}\right)^{p_1} + \left(\frac{1+z_c}{1+z} \right)^{p_2}}
\end{equation}
and
\begin{equation}
    e_2(z) = (1+z)^{k_1}
\end{equation}
with $\log_{10}\phi_1 = -3.749$, $\log_{10}L_* = 21.592$, $\beta = 0.139$, $\gamma = 0.878$, $z_c = 0.893$, $p_1 = 2.085$, $p_2 = -4.602$, and $k_1 = 1.744$.

\begin{figure}
    \centering
    \includegraphics[width = \linewidth]{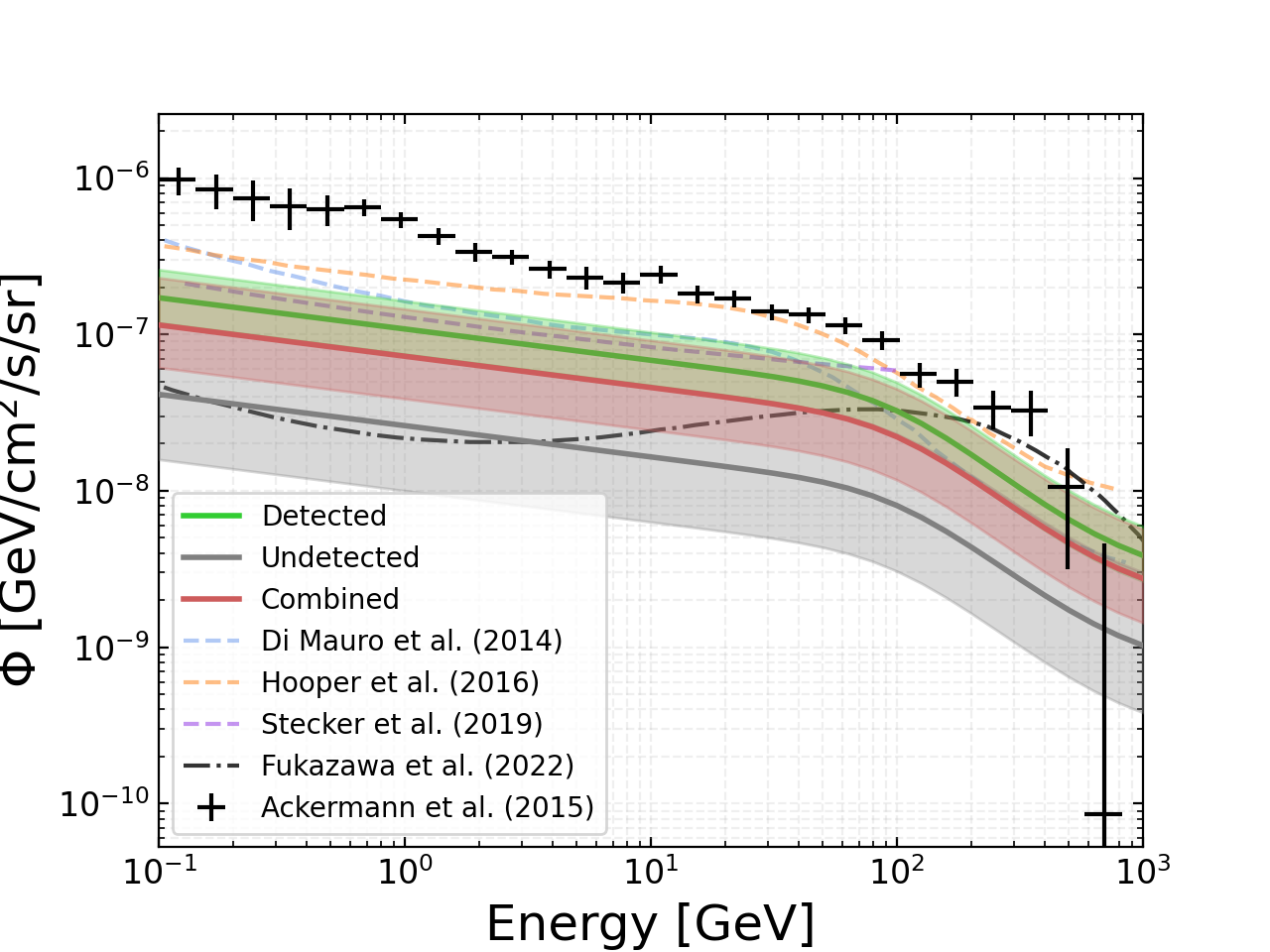}
    \caption{Integrated $\gamma$-ray emission from RGs.
    The green band is the predicted contribution from the detected RG population, with its $1\sigma$ uncertainty.
    The gray band is the predicted contribution from the subthreshold RG population with its $1\sigma$ uncertainty.
    The red band is the contribution obtained combining the results of both samples of RGs, with its $1\sigma$ uncertainty. The dashed blue, orange and purple lines are the contributions from \cite{DiMauro+2014}, \cite{Hooper+2016}, and \cite{Stecker+2019}, respectively.
    The black dot-dashed line is the contribution from \cite{Fukazawa+2022}.
    The black datapoints are the LAT evaluation of the IGRB \citep[see][]{Ackermann2015}}
    \label{fig:EGB}
\end{figure}

Once the GLF has been determined, the contribution to the IGRB from a population of \gray sources can be evaluated as
\begin{align}
\begin{split}
    \Phi(\epsilon) = & \int_{0}^{z_{max}} dz \frac{d^2 V}{dz d\Omega} \int_{\log_{10}(L_{\gamma,min})}^{\log_{10}(L_{\gamma,max})} d(\log_{10}L_\gamma) \frac{dF_\gamma}{d\epsilon}\\
    & \left(1 - \omega(F_{\gamma}(L_{\gamma}, z))\right) e^{-\tau_{\gamma\gamma}(\epsilon, z)} \rho_\gamma(L_\gamma, z).
\end{split}
\end{align}
We compute the contribution from our samples separately, due to the differences in the parameters shown in the previous sections.
For both computations we use the same integration limits. 
The minimum \gray luminosity is $L_{\gamma, min} = 10^{38}\, \textrm{erg s}^{-1}$, while the maximum luminosity is $L_{\gamma, max} = 10^{46}\, \textrm{erg s}^{-1}$. 
The maximum redshift is $z = 3.5$, to represent the redshift range of the RGs observed in \citet{YuanWang2012}.
We use the upper bounds of the combined samples as integration limits, as the IGRB computation is less sensitive to the upper bounds for $L_\gamma$ and $z$, due to the shape of the GLF.
For a complete derivation, see \citet{Ackermann2015}.
The factor $dF_\gamma/d\epsilon$ is the intrinsic photon flux at energy $\epsilon$ for a source with \gray luminosity $L_\gamma$ and redshift $z$.
For both subthreshold and individual sources we assume a power law with index $\Gamma = 2.4^{+0.2}_{-0.2}$ and $\Gamma = 2.2^{+0.1}_{-0.2}$, respectively.
The index for subthreshold RGs is derived in \S\ref{sec:analysis}.
The index for detected RGs is evaluated through the same stacking method, summing the individual TS profiles in the flux -- index space.
The term $\omega(F_{\gamma}(L_{\gamma}, z))$ is the flux-dependent detection efficiency of the \fermi-LAT at given \gray luminosity and redshift.
The values of $\omega(F_{\gamma}(L_{\gamma}, z))$ are interpolated from \cite{Marcotulli+2020}.
During their propagation towards Earth, the high energy photons emitted by the source ($\epsilon > 20 \gev$) can interact with the extragalactic background light (EBL), through absorption or pair production \citep{Gould+1966, Stecker+1992, Finke+2010}.
We use the model from \citet{Finke+2010} to describe the resulting absorption effect, where $\tau_{\gamma\gamma}(\epsilon, z)$ is the optical depth.
In pair production, the interaction of the $\gamma$-rays with the EBL generates electron-positron pairs, which can then produce a cascade emission at lower \gray energies, via inverse Compton scattering of cosmic microwave background photons. 
However, this effect is small for source populations with a photon index greater than 2 \citep{Inoue+2012}, which is the case for both of our samples. 
We therefore neglect the cascade emission in our analysis.
The last factor in the integral is the comoving volume element, defined in the standard flat $\Lambda$CDM cosmology as
\begin{equation}
    \frac{d^2 V}{dz d\Omega} = \frac{c d_L^2(z)}{H_0(1+z)^2\sqrt{\Omega_{M,0}(1+z)^3 + \Omega_{\Lambda,0}}}
\end{equation}
where $d_L(z)$ is the luminosity distance at redshift $z$.
In Fig.~\ref{fig:EGB} we compare the result for the detected sample with the contributions evaluated in \citet{DiMauro+2014}, \citet{Hooper+2016}, \citet{Stecker+2019}, though only the latter uses our same RLF, while the first two use the RLF from \citet{YuanWang2012}.
The prediction from the sample of detected RGs shows good agreement with previous results. 
The best fit line for this contribution is $\sim 31\%$ of the IGRB measured by the LAT \citep{Ackermann2015}, while its $1\sigma$ uncertainty is between $17\%$ and $63\%$ -- determined, as all the uncertainties hereafter, from the uncertainties on the parameters $\alpha$ and $\beta$ of the $\LL$ correlation.
The best-fit line for the contribution from the sample of subthreshold RGs is $\sim 8\%$ of the IGRB, with its uncertainty within $4\%$ and $17\%$.
These two predictions act as upper and lower bounds for the contribution from the full population of RGs.
Using the parameters of the $\LL$ correlation obtained from the combination of the two samples, we can derive an estimation of the contribution to the IGRB that is representative of the entire population of RGs. 
This contribution is $\sim 21\%$ of the IGRB estimated by LAT, with its $1\sigma$ uncertainty between $\sim 11\%$ and $\sim 44\%$.
This result, which is the major finding of the present paper, predicts a diffuse \gray emission from RGs a bit smaller than previously determined. The very difference with state-of-the-art literature is in the inclusion of subthreshold sources.

\section{Source Count Distribution}
\label{sec:scd}
\begin{figure}
    \centering
    \includegraphics[width = \linewidth]{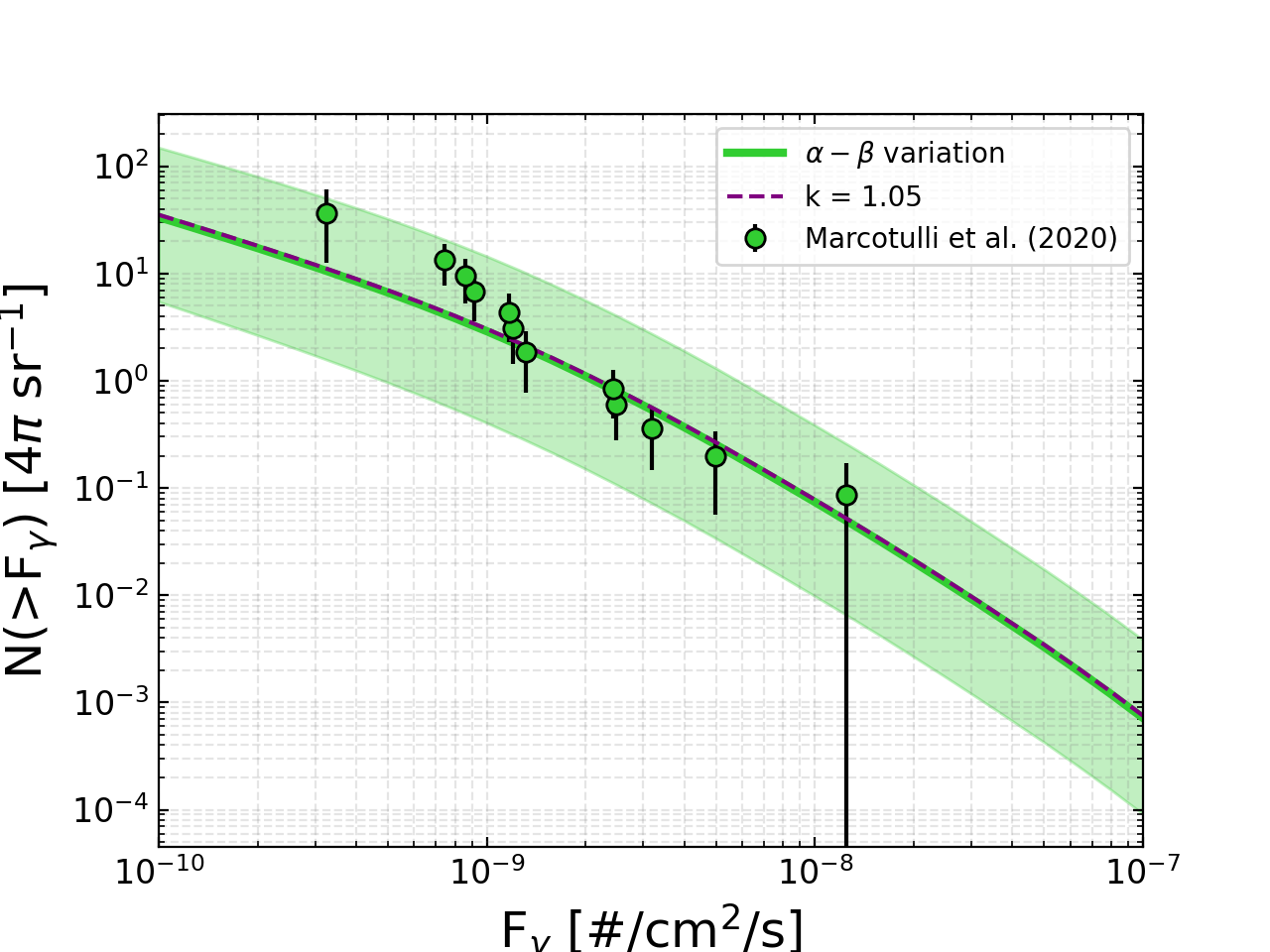}
    \caption{Source count distribution for the RGs. 
    The data points are obtained using the efficiency and flux values of the sources of the detected sample that are in \cite{Marcotulli+2020}. 
    The dashed red line is obtained by fitting the source-count distribution obtained assuming the best-fit parameters for the $L_\gamma - L_{C,5\textrm{GHz}}$ correlation from the detected RGs to the data points.
    The green band is obtained fixing the parameter $k = 1$, and variating the $\alpha$ and $\beta$ parameters from the detected RGs, within their $1\sigma$ uncertainties.}
    \label{fig:scd}
\end{figure}

In evaluating the proportionality between the GLFs and RLFs, the parameter $k$ (see Eq. \ref{eq:GLF}) plays a fundamental role.
This parameter expresses the fraction of \gray emitting sources in the population that also have a radio-loud core:
\begin{equation}
    N_\gamma = k N_R.
\end{equation}
The value $k = 1$ corresponds to the hypothesis that every RG emitting in $\gamma$-rays also has an emission from the radio core, and allows to compute the GLF from the RLF only using the $L_\gamma - L_{C,5\textrm{GHz}}$ correlation.
While a definitive value for $k$ is difficult to obtain with the incomplete information we have about the population, we can estimate its order of magnitude through the study of the source-count distribution, $N(>F_\gamma)$.
The theoretical source-count distribution is defined, at the \gray energies, as follows:
\begin{align}
    \nonumber N(>F_\gamma) =& \;4\pi \int^{\Gamma_{min}}_{\Gamma_{max}} \frac{dN}{d\Gamma} d\Gamma
    \int_{0}^{z_{max}} \frac{d^2V}{dzd\Omega}dz\\
    &\times \int_{L_{\gamma}(F_\gamma, z, \Gamma)}^{L_{\gamma}^{max}} \frac{dL_\gamma}{L_{\gamma} \ln(10)} \rho_{\gamma}
    (L_\gamma, z, \Gamma),
\end{align}
where $\rho_\gamma$ is the GLF from Eq.~\ref{eq:GLF}. $L_{\gamma}(F_\gamma, z, \Gamma)$ is the \gray luminosity of a RG with photon flux $F_\gamma$ and photon index $\Gamma$, at redshift $z$.
The spectral index distribution, $dN/d\Gamma$, is here taken to be a Dirac delta $\delta(\Gamma - \Gamma_0)$ at $\Gamma_0 = 2.3$. 
While \fermi-LAT, as any other observing instrument, inevitably select sources closer to its detection significance, leading to an asymmetrical observed photon index distribution, the observed RGs span a range of photon indexes too narrow to have strong effects on the computation of the integral.
The upper limits of the integration are $z_{max} = 3.5$ and $L_{\gamma}^{max} = 10^{50} \mathrm{erg\;s}^{-1}$.
 Integrating to higher values of $L_{\gamma}^{max}$ and $z_{max}$ would have a marginal effect on the computation, due to the shape of the GLF.\\
To obtain an estimation of the parameter $k$, we define an experimental source-count distribution from the RGs detected by the \fermi-LAT, as:
\begin{equation}
    N(>F_\gamma) = \sum^{N(>F_{\gamma,i})}_{i = 1} \frac{1}{\omega(F_{\gamma,i})}.
\end{equation}
The inclusion of $\omega(F_{\gamma,i})$ comes from the fact that, as we mentioned above, \fermi-LAT more easily detects soft-spectrum sources at faint fluxes, skewing the observed photon-index distribution.
The detection efficiency used here is from \cite{Marcotulli+2020}. 
For consistency, we only evaluate the source-count distribution points using the 11 sources in our detected sample that are listed in the catalog presented by \cite{Marcotulli+2020}, with the coordinates within their 95\% confidence radii.
We then fit the theoretical source-count distribution to the data points constructed this way, to evaluate $k$.
Since the experimental points are highly correlated, the uncertainty of $k$ obtained from the fit is not statistically significant. 
Yet, evaluating $k$ through this procedure is still useful to confirm that it is of the order of the unit.
From the fit, we obtain $k = 1.05 \pm 0.10$ (see Fig.~\ref{fig:scd}) with a $\chi^2 = 9.51$ (for 11 degrees of freedom), which confirms the hypothesis made in Sec~\ref{sec:gamma_radio_corr} when defining the GLF.
Even though the fit on the parameter $k$ tends to underestimate its uncertainty, this is negligible when compared to the uncertainty coming from the determination of the parameters of the $\LL$ correlation.
In Fig.~\ref{fig:scd}, we display this effect by fixing $k = 1$ and computing the source-count distribution for values of $\alpha$ and $\beta$ within their $1\sigma$ uncertainties from the analysis of the detected sample (i.e., $\alpha = 1.0^{+0.1}_{-0.1}$ and $\beta = 44.0^{+0.1}_{-0.1}$).

\section{Discussion and conclusions}
\label{sec:disc}
\noindent We have used 15.9 years of \textit{Fermi}-LAT data to characterize the \gray emission from RGs and their contribution to the IGRB.
We carried out our analysis using two separate samples of RGs: one containing the few previously detected by the LAT and included in the 4FGL-DR3 catalog, and a separate sample containing RGs detected at radio frequencies, but lacking significant \gray detection.
Through the use of a stacking technique the emission from the subthreshold population is detected at a TS value of $67.8$  corresponding to a significance of  $7.9 \sigma$.
The average SED, assuming a power-law behaviour, was constrained to have a photon index $\Gamma = 2.4^{+0.2}_{-0.2}$.
Similarly, the average SED of the detected RGs was found to have a photon index $\Gamma = 2.2^{+0.1}_{-0.2}$.

The \gray emission for both sets of RGs is correlated to the radio emission from the core at 5 GHz.
For the detected sample, the parameters of the correlation were evaluated, through the stacking procedure, to be $\alpha = 1.0^{+0.1}_{-0.1}$ and $\beta = 44.0^{+0.1}_{-0.1}$, with a peak TS value of $TS = 3488$, corresponding to a significance of $\sim 59 \sigma$.
The parameters of the subthreshold sample, also obtained from the stacking, are $\alpha = 0.6^{+0.1}_{-0.1}$ and $\beta = 42.6^{+0.1}_{-0.2}$, reaching a TS value of $TS = 87.8$, corresponding to a significance of $\sim 9.1\sigma$.
The parameters of this correlation differ between the two samples, implying that for the same 5 GHz luminosities, sub-threshold RGs are fainter than detected ones.
In $\S$ \ref{sec:stat_test}, we showed that this is likely not a bias due to including sources that are too faint even for the stacking analysis.
We then combined the results from the two separate samples, obtaining correlation parameters that are representative of the entire population of RGs.
These were found to be $\alpha = 0.7^{+0.1}_{-0.1}$ and $\beta = 43.3^{+0.1}_{-0.1}$, with a peak TS of $TS = 4058$, corresponding to a significance of $\sim 63 \sigma$.

Finally, we computed the contribution from the RG population to the IGRB using the average spectral index of the detected sample, along with its values for the  $L_\gamma - L_{C,5\textrm{GHz}}$ correlation parameters, and then separately using the corresponding values from the subthreshold sample, and the ones from the combination of the two samples. 
We compared our results to the IGRB measured by the LAT \citep{Ackermann2015}.
The best fit for the contribution to the IGRB from the detected sample is at $\sim 31\%$ of the LAT estimate.
When considering the $1\sigma$ uncertainty, evaluated from the uncertainty on the parameters $\alpha$ and $\beta$ of the $\LL$ correlation, this contribution can be as low as $\sim17\%$ and as high as $\sim63\%$.
For the subthreshold sample, the best fit for the contribution to the IGRB is at $\sim 8\%$ of the LAT estimate. 
When including the $1\sigma$ uncertainty, it can go down to $\sim4\%$ and up to $\sim 17\%$.
It is likely that these two bands represent extreme cases that bracket the values of the contribution to the IGRB coming from the RG population as a whole, which is best represented by the estimation obtained using the correlation parameters from the combination of the two samples.
This kind of combined contribution is at $\sim 21 \%$ of the LAT estimate, with its $1\sigma$ uncertainty between $\sim 11\%$ and $\sim 44\%$. 

In Fig.~\ref{fig:EGB}, we compare our results with the IGRB data measured by the LAT, and previous determinations of the RG contribution from \citet{DiMauro+2014}, \citet{Hooper+2016}, \citet{Stecker+2019}, and \cite{Fukazawa+2022}.
Our results show reasonable agreement with the previous determinations, although preferring lower values of the integrated emission compared to \citet{DiMauro+2014} and \citet{Hooper+2016}, which use an older model for the RLF, based on the total radio emission of the RGs, rather than the emission of the radio cores.
The line from \citet{Stecker+2019}, whose data sample is the most similar to our detected sample and uses our same RLF, shows the best agreement with our determination.
The biggest addition of our analysis to these studies, however, is the determination of the contribution using parameters evaluated from the combination of the detected and subthreshold populations of RGs, which represent a more realistic estimation of the diffuse emission coming from the whole population of RGs.
Furthermore, the results should be compared to the contributions evaluated for other classes of sources. Blazars can contribute to $\sim 30\%$ of the IGRB \citep{Ajello2015, Korsmeier+22}, while the contribution from star-forming galaxies can range from a small percentage up to the totality of the background \citep{Ajello+2020a, Roth+21}. Looking at these results together rather than individually can help to constrain the large uncertainties that affect them, and give a more complete view of the composition of the IGRB.

\section*{Acknoledgements}
\noindent The \textit{Fermi}-LAT Collaboration acknowledges generous ongoing support from a number of agencies and institutes that have supported both the development and the operation of the LAT as well as scientific data analysis.
These include the National Aeronautics and Space Administration and the Department of Energy in the United States, the Commisariat à l'Energie Atomique and the Centre National de la Recherche Scientifique / Institut National de Physique Nucléaire et de Physique des Particules in France, the Agenzia Spaziale Italiana and the Istituto Nazionale di Fisica Nucleare in Italy, the Ministry of Education, Culture, Sports, Science and Technology (MEXT), High Energy Accelerator Research Organization (KEK) and Japan Aerospace Exploration Agency (JAXA) in Japan, and the K. A. Wallenberg Foundation, the Swedish Research Council and the Swedish National Space Board in Sweden.

Additional support for science analysis during the operations phase is gratefully acknowledged from the Istituto Nazionale di Astrofisica in Italy and the Centre National d'\`Etudes Spatiales in France.
This work performed in part under DOE Contract DE-AC02-76SF00515.

This research has made use of the NASA/IPAC Extragalactic Database (NED), which is operated by the Jet Propulsion Laboratory, California Insitute of Technology, under contract with the National Aeronautics and Space Administration. 

Clemson University is acknowledged for their generous allotment of compute time on the Palmetto Cluster.

C.M.K.’s research was supported by an appointment to the NASA Postdoctoral Program at NASA Goddard Space Flight Center, administered by Oak Ridge Associated Universities under contract with NASA.

M.D.M. and  F.D. acknowledge the support of the Research grant {\sc TAsP} (Theoretical Astroparticle Physics) funded by Istituto Nazionale di Fisica Nucleare. 
F.D. acknowledges the Research grant {\sl Addressing systematic uncertainties in searches for dark matter}, Grant No. 2022F2843L funded by the Italian Ministry of Education, University and Research (MIUR).

The NASA/IPAC Extragalactic Database (NED) is funded by the National Aeronautics and Space Administration and operated by the California Institute of Technology.

\bibliography{ref.bib}

\end{document}